\documentclass{JINST}
\usepackage{epsfig}

\usepackage{textcomp}
\title{Long Term Performance Studies of Large Oil-Free Bakelite Resistive Plate Chamber}

\author{Rajesh Ganai$^a$\thanks{Corresponding author.}, Arindam Roy$^a$, Mehul Kumar Shiroya$^b$, Kshitij Agarwal$^c$, 
Zubayer Ahammed$^a$, Subikash Choudhury$^a$ and Subhasis Chattopadhyay$^a$\\
\llap{$^a$}Variable Energy Cyclotron Centre,\\
  1/AF-Bidhan Nagar, Kolkata-700064, India\\
\llap{$^b$}Sardar Vallabhbhai National Institute of Technology,\\
  Surat, Gujarat-395007, India\\
\llap{$^c$}Birla Institute of Technology and Science,\\
  Pilani, Rajasthan-333031, India\\
  E-mail: \email{rajesh.ganai.physics@gmail.com}}

\abstract{Several high energy physics and neutrino physics experiments worldwide require large-size RPCs to cover wide acceptances. 
The muon tracking systems in the Iron calorimeter (ICAL) in the INO experiment, India and the near detector in DUNE at Fermilab are
two such examples. A (240 cm $\times$ 120 cm $\times$ 0.2 cm) bakelite RPC has been built and tested at Variable Energy Cyclotron 
Centre, Kolkata, using indigenous materials procured from the local market. No additional lubricant, like oil has been used on the 
electrode surfaces for smoothening. The chamber is in operation for $>$ 365 days. We have tested the chamber for its long term operation. 
The leakage current, bulk resistivity, efficiency, noise rate and time resolution of the chamber have been found to be quite stable 
during the testing peroid. It showed an efficiency $>$ 95$\%$ with an average time resolution of $\sim$0.83 ns at the point of measurement
at 9000 V throughout the testing period. Details of the long term performance of the chamber have been discussed.}


\begin{document}

\section{Introduction}

Resistive Plate Chambers (RPCs) \cite{santonico_cardarelli} will find possible use to track muons in the 
the upcoming neutrino experiments like Iron CALorimeter (ICAL) in India based Neutrino Observatory (INO) \cite{ino} in India, 
Deep Underground Neutrino Experiment (DUNE) \cite{dune} at Fermilab,USA. 
INO-ICAL of dimension $\sim$ 48 m $\times$ 16 m $\times$ 14 m will consist of $\sim$ 50 kT magnetised iron plates stacked in 150 
layers. About 30,000 single gap RPC modules each of dimension $\sim$ 200 cm $\times$ 200 cm $\times$ 0.2 cm 
sandwiched between pairs of 150 layers
of iron plates will be used as tracking layers. The RPC modules to be used in  DUNE are of dimension 200 cm $\times$ 100 cm $\times$ 0.2 cm.
RPCs, because of very good time resolution, will be capable of differentiating between up and down going neutrinos in INO-ICAL.\\
As a part of this programme, Variable Energy Cyclotron Centre (VECC), Kolkata, India, has been actively involved the R$\&$D of RPCs 
using high pressure paper laminates, commonly referred as bakelite, for almost a decade. VECC aims in the development of 
large size (240 cm $\times$ 120 cm) RPCs that might be used in 
INO-ICAL or in the Near Detector (ND) of the Deep Underground Neutrino Experiment (DUNE). 
As these experiments will run for many years, therefore it is necessary that the stability of this RPC in long term operation is tested.\\
Long term operation of a RPC may cause ageing problems in the detector due to several reasons like the variation of the electrode
resistivity, the integrated charge etc. 
In general there are three reasons of ageing effects in a RPC- ageing of the electrode material, aging due to the integrated charge generated
in the gas gap which results in the increase in current in the detector and
%integrated dissipated current  the detector,
ageing due to irradiation on the detector. However, for low rate neutrino experiments like INO, ageing due to irradiation might not be
significant but ageing due to the others remain major issue.
Increase in the leakage current, bulk resistivity of the chamber, reduction in efficiency are the prominent symptoms of an aged RPC.
It is therefore important to perfom the long term test of a detector.\\
In this paper, we report the long term performance studies of a (240 cm $\times$ 120 cm $\times$ 0.2 cm) RPC 
that uses 3 mm thick bakelite sheets as electrodes.

\section{Fabrication of large RPC}
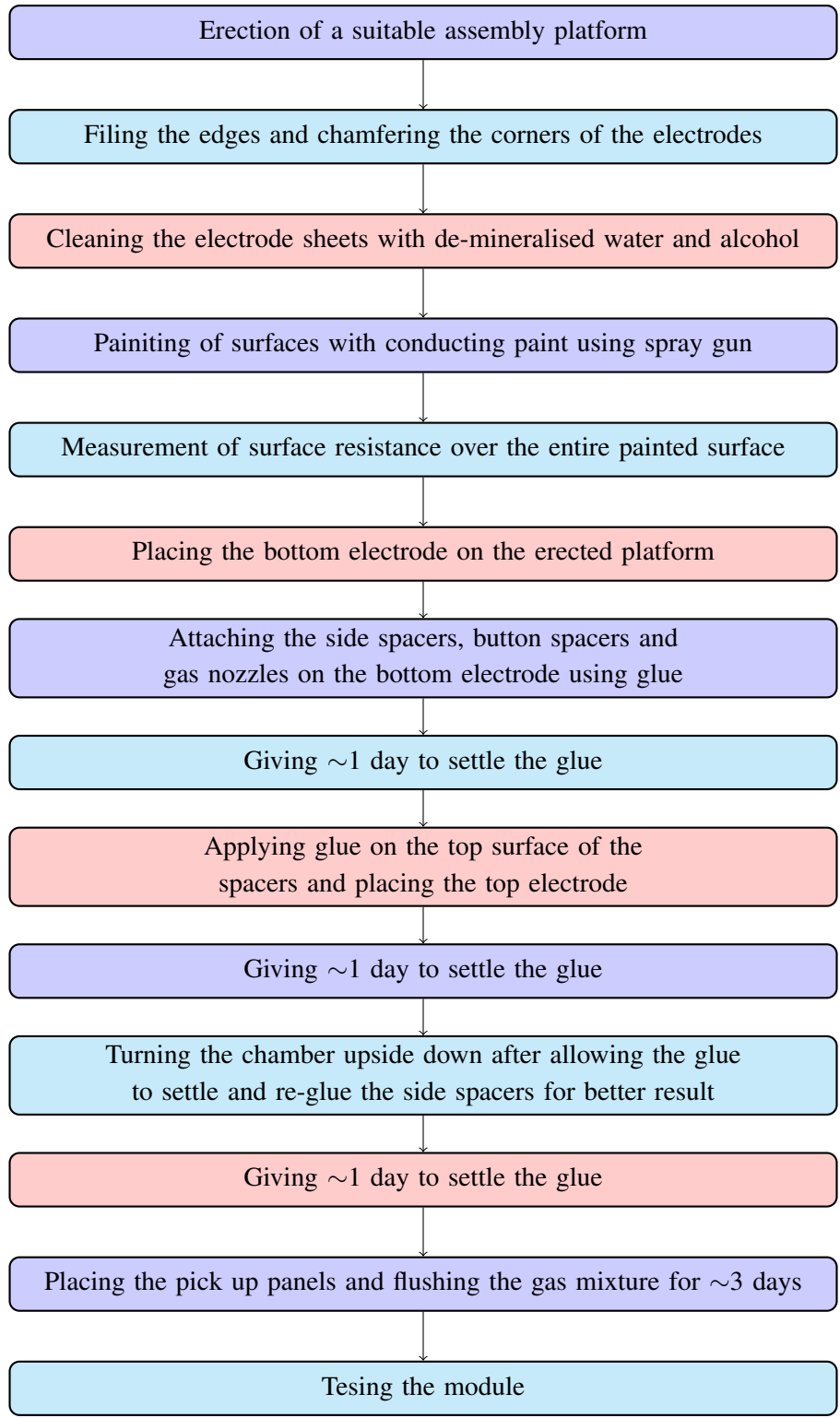
\begin{figure}
\begin{center}
\begin{tikzpicture}[  
  decision/.style = { rectangle, draw=blue, thick, fill=blue!20, text width=5em, text badly centered, inner sep=2pt, rounded corners },
  block1/.style    = { rectangle, draw=black, thick, fill=blue!20, text width=30em, text centered, rounded corners, minimum height=2em },
  block2/.style    = { rectangle, draw=black, thick, fill=cyan!20, text width=30em, text centered, rounded corners, minimum height=2em },
  block3/.style    = { rectangle, draw=black, thick, fill=red!20, text width=30em, text centered, rounded corners, minimum height=2em },
  %block2/.style    = { rectangle, draw=black, thick, fill=red!20, text width=30em, text centered, rounded corners, minimum height=4em },
  line/.style     = {arrow, thick, -> },]

\node [block1] (step1) {Erection of a suitable assembly platform};

\node [block2, below of=step1, yshift=-0.5cm] (step2) {Filing the edges and chamfering the corners of the electrodes};

\node [block3, below of=step2, yshift=-0.5cm] (step3) {Cleaning the electrode sheets with de-mineralised water and alcohol};

\node [block1, below of=step3, yshift=-0.5cm] (step4) {Painiting of surfaces with conducting paint using spray gun};

\node [block2, below of=step4, yshift=-0.5cm] (step5) {Measurement of surface resistance over the entire painted surface};

\node [block3, below of=step5, yshift=-0.5cm] (step6) {Placing the bottom electrode on the erected platform};

\node [block1, below of=step6, yshift=-0.5cm] (step7) {Attaching the side spacers, button spacers and gas nozzles on the bottom electrode 
							using glue};

\node [block2, below of=step7, yshift=-0.5cm] (step8) {Giving $\sim$1 day to settle the glue};

\node [block3, below of=step8, yshift=-0.5cm] (step9) {Applying glue on the top surface of the spacers and placing the top electrode};

\node [block1, below of=step9, yshift=-0.5cm] (step10) {Giving $\sim$1 day to settle the glue};

\node [block2,  below of=step10, yshift=-0.5cm] (step11) {Turning the chamber upside down after allowing the glue to settle and re-glue
							  the side spacers for better result};
							  
\node [block3, below of=step11, yshift=-0.5cm] (step12) {Giving $\sim$1 day to settle the glue};

\node [block1, below of=step12, yshift=-0.5cm] (step13) {Placing the pick up panels and flushing the gas mixture for $\sim$3 days};

\node [block2, below of=step13, yshift=-0.5cm] (step14) {Tesing the module};

	\draw [->] (step1) -- (step2);
	\draw [->] (step2) -- (step3);
	\draw [->] (step3) -- (step4);
	\draw [->] (step4) -- (step5);
	\draw [->] (step5) -- (step6);
	\draw [->] (step6) -- (step7);
	\draw [->] (step7) -- (step8);
	\draw [->] (step8) -- (step9);
	\draw [->] (step9) -- (step10);
	\draw [->] (step10) -- (step11);
	\draw [->] (step11) -- (step12);
	\draw [->] (step12) -- (step13);
	\draw [->] (step13) -- (step14);

\end{tikzpicture}
\caption{\small \sl Flow-chart of steps followed during fabricating of large RPC}
\label{flow-chart}
\end{center}
\end{figure}

\begin{figure}
\centering
%\sidecaption
\includegraphics[height=10.0cm,width=15.0cm]{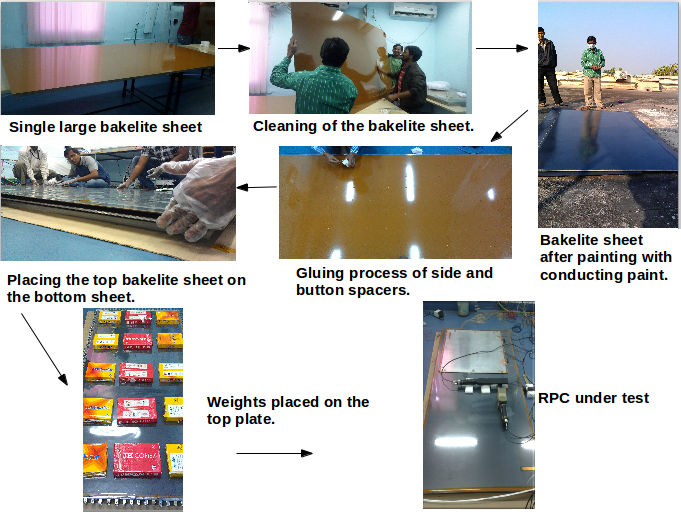}
\caption{[Color online] \small \sl Photographs of the steps followed during fabricating the large bakelite RPC\cite{small_RPC}.}
\label{large_fab_steps} 
\end{figure}

\begin{figure}
\begin{center}
\includegraphics[height=4.2 cm, width=5.6 cm]{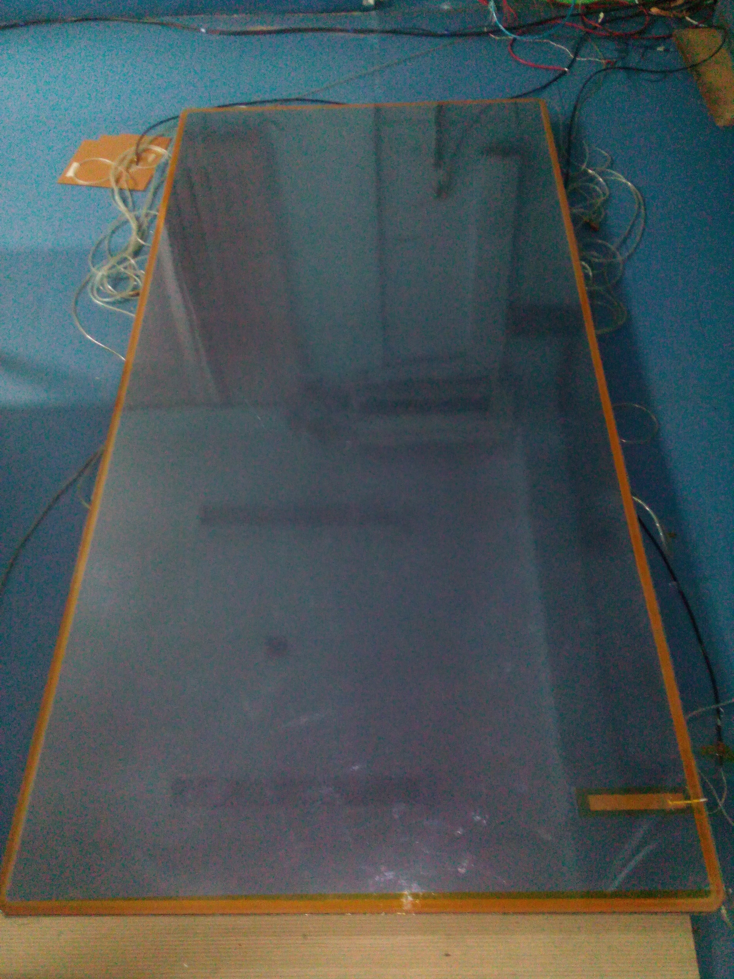}
\caption{\label{setup2} \small \sl [Color online] Photograph of the large bakelite RPC\cite{large_RPC_VECC}.}
\label{full-rpc}
\end{center}
\end{figure}

\begin{figure}
\begin{center}
\includegraphics[height=4.0 cm, width=7.0 cm,keepaspectratio]{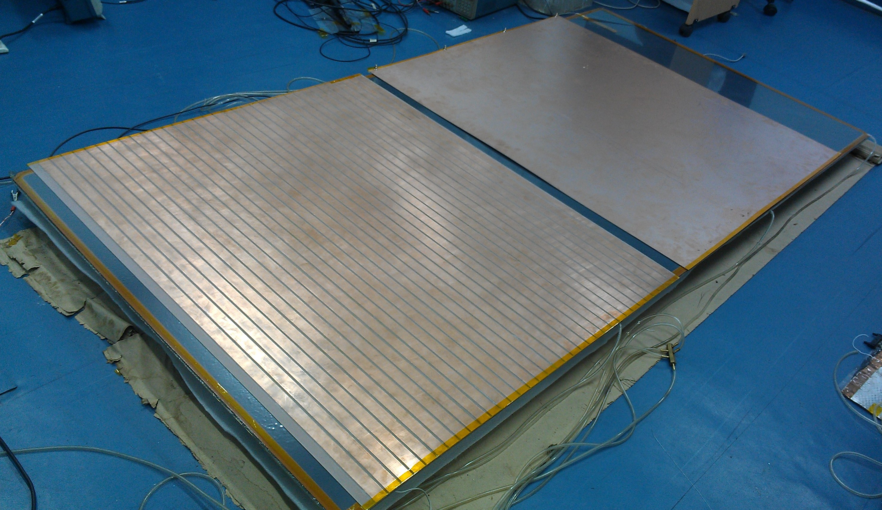}
\caption{\label{setup2} \small \sl [Color online] Photograph of the large bakelite RPC with pick-up panel\cite{large_RPC_VECC}.}
\label{full-rpc-with-pickup}
\end{center}
\end{figure}

We have successfully fabricated a $\sim$ 240 cm $\times$ 120 cm $\times$ 0.2 cm, oil-free bakelite RPC using raw materials like bakelite
sheets, glue, spacers available from Indian market. Several grades of bakelites available in Indian market have been tried out for
successfull development of oil-free RPC. Successfull fabrication and results of \cite{small_RPC} encouraged us to fabricate the large RPC.
For completeness, we give a brief account of the fabrication procedure. Fig.~\ref{flow-chart} and Fig.~\ref{large_fab_steps}
show various steps involved and a set of photographs taken while fabricating the large bakelite RPC. 
Many challenges were faced during fabricating such a large size RPC. The most crucial challenge was to maintain the planarity of the
platform on which the RPC was fabricated. For this, a suitable platform was made with the help of cardboard sheets, foams, paper bundles and
thick glass plate. The next challenge was to select a suitable glue to stick the gas nozzles, side and button spacers. A detailed
R$\&$D on different glue samples was done and a suitable glue was chosen. The number of gas nozzles to be used, the number and dimension of button 
spacers and side spacers were also decided after detailed study. 
The details of the fabrication process and test results using cosmic rays have been
discussed in \cite{large_RPC_VECC}. Fig.~\ref{full-rpc} and Fig.~\ref{full-rpc-with-pickup} show the fabricated RPC.
In order to tap the signals from the chamber, pick up pannles which are made of 
$\sim$(125 cm $\times$ 105 cm $\times$ 0.15 cm) FR4 sheet sandwiched between $\sim$(125 cm $\times$ 105 cm $\times$ 0.0035 cm) 
copper sheets have been used. The copper pick-up strips are 2.5 cm in width, with a gap of 0.2 cm 
between adjacent strips.

% Define block styles

\section{Long term test results using cosmic rays and discussions}
A scintillator based cosmic ray test set up has been used for the test. During the entire testing period, the laboratory temperature
and the relative humidity have been maintained at $\sim$ 10\textdegree C and $\sim$35\% - 40\% respectively. 
All the tests have been performed in the streamer mode of operation of the RPC with a gas composition of 
Argon:Freon(R134a):Iso-butane::34:57:9 by volume. 
A typical gas flow rate of $\sim$0.75 litre/hour has been maintained over the entire test period resulting in $\sim$3 changes
of gas volume per day. The chamber was operated at 9000 V with a signal threshold of -20 mV.\\
Positive and negative high voltage to the chamber was applied on the top and bottom electrodes with CAEN A1832PE and A1832NE modules 
respectively, in a CAEN SY1527 crate.
The current was monitored from the panel of the HV supply. LEMO connectors were soldered on the copper strips of the pick-up panel
to tap the signals from the chamber. The raw RPC signal was digitised using a CANBERRA QUAD CFD 454 constant fraction
discriminator (CFD). 
A CAMAC based data acquisition system has been used in our setup. 
For timing measurements, a 16 channel PHILIPS SCIENTIFIC 7186 TDC module was used.
The master trigger was formed with one finger (7 cm $\times$ 1.5 cm) and two paddle (20 cm $\times$ 8.5 cm) scintillators.
The overlap area between the scintillators has been used to obtain the cosmic ray efficiency.
The average master trigger rate was $\sim$ 0.0085 Hz/$cm^2$.\\
The chamber is in gas and high voltage for more than a year. We report here the monitoring of chamber current,
efficiency, noise rate and time resolution of the chamber continuously for 60 days. Data for chamber current, efficiency and noise rate
have been taken every day whereas the time resolution measurement has been done once in a week's time.

\subsection{Current stability}
Fig.~\ref{current_stab} shows the variation in current in both the electrodes. The top and middle figures show the 
current variation for the top and bottom electrodes respectively whereas the bottom figure shows the average current 
of both the electrodes. The current in both the electrodes was found to be quite stable during the monitoring period.
We have calculated the bulk resistivity of the chamber from the average current values which is shown in Fig.~\ref{bulk_with_time}.
The bulk resistivity of the chamber was found to be $\sim$ 9 $\times$ $10^{13}$ $\Omega$cm.

\begin{figure}
\begin{center}
\includegraphics[height=5.0 cm, width=8.0 cm]{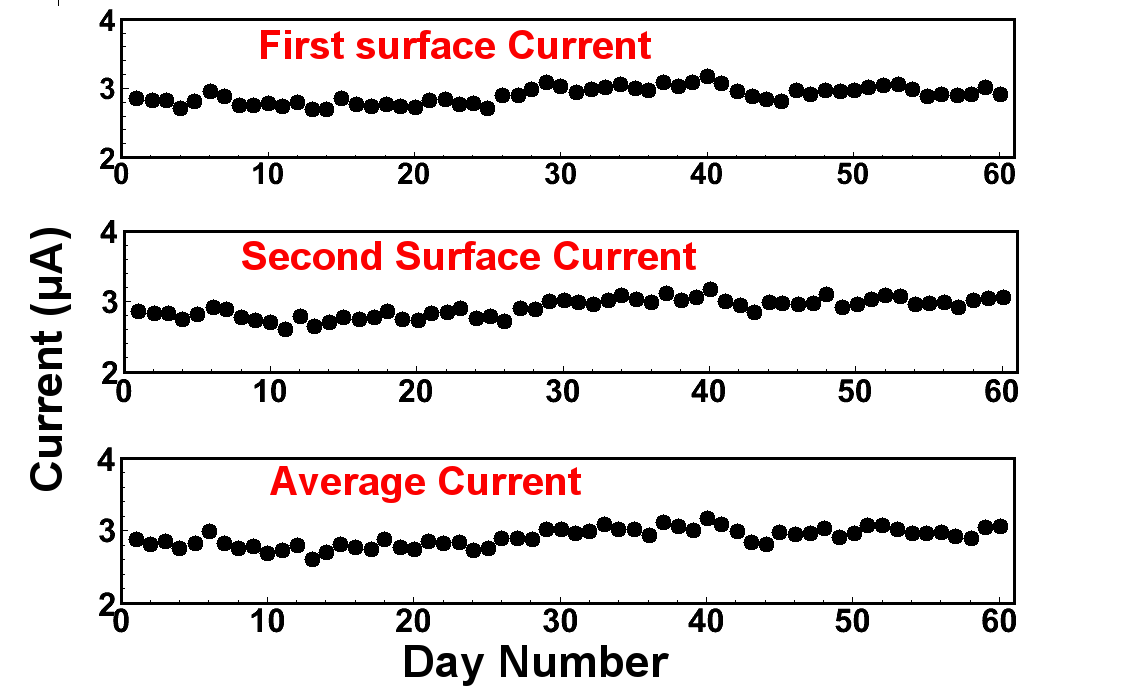}
\caption{\label{current_stab} \small \sl Current stability of the electrodes of the chamber over a peroid of 60 days.}
\label{current_stab} 
\end{center}
\end{figure}

\begin{figure}
\begin{center}
\includegraphics[height=5.0 cm, width=8.0 cm]{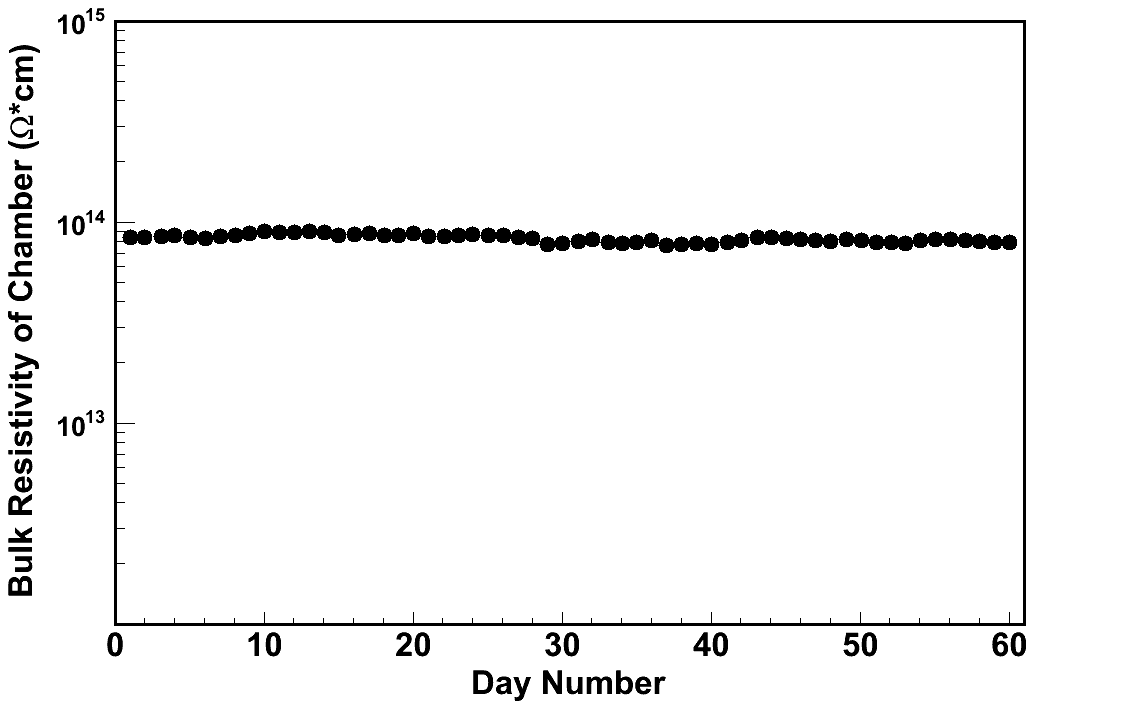}
\caption{\label{bulk_with_time} \small \sl [Color online] Variaton of bulk resistivity of the chamber with time.}
\label{bulk_with_time} 
\end{center}
\end{figure}

\subsection{Efficiency and noise rate}
The values of efficiency and noise rate have been shown
in Fig.~\ref {efficiency_time} and Fig.~\ref {noise_rate_time} respectively.
The average values of efficiency and noise rate of the chamber were found to be $>$95$\%$ and $\sim$ 0.75 Hz/$cm^{2}$ respectively. 
The graphs showed that both remained quite stable over the entire testing period.

\begin{figure}
\begin{center}
\includegraphics[height=5.0 cm, width=8.0 cm]{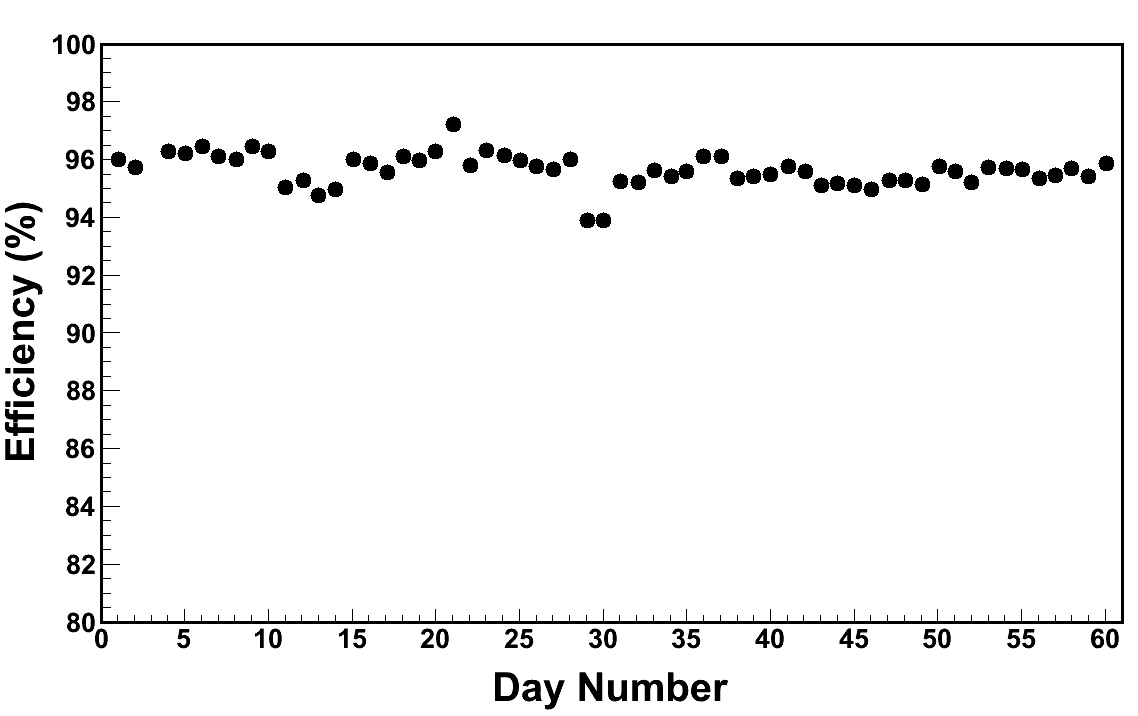}
\caption{\label{efficiency_time} \small \sl Efficiency of the chamber over a period of 60 days. The error bars are within the marker size.}
\label{efficiency_time}
\end{center}
\end{figure}

\begin{figure}
\begin{center}
\includegraphics[height=5.0 cm, width=8.0 cm]{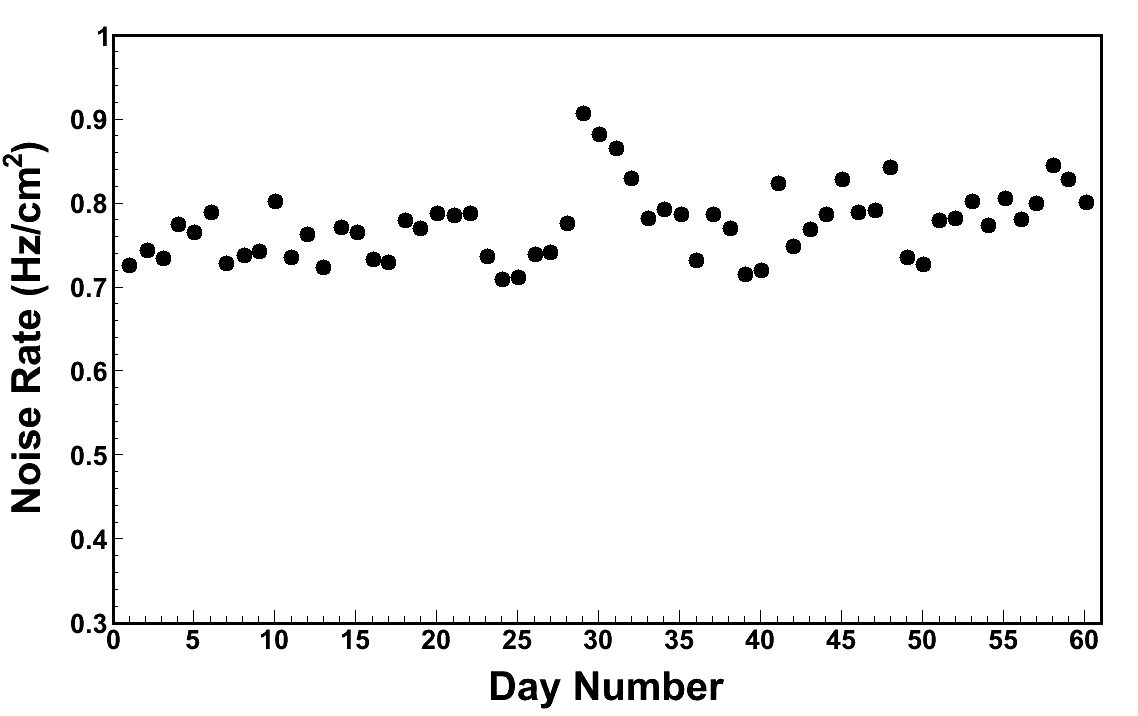}
\caption{\label{noise_rate_time} \small \sl [Color online] Noise rate of the chamber over a period of 60 days. The error bars are 
within the marker size.}
\label{noise_rate_time}
\end{center}
\end{figure}

\subsection{Time resolution}
Fig.~\ref {time-resolution_time} shows the variation of time resolution ($\sigma$) over the entire testing period.
The details of the time resolution ($\sigma$) calculations have been discussed in \cite{large_RPC_VECC}. The average
time resolution of the chamber was found to be $\sim$ 0.83 ns.

\begin{figure}
\begin{center}
\includegraphics[height=5.0 cm, width=8.0 cm]{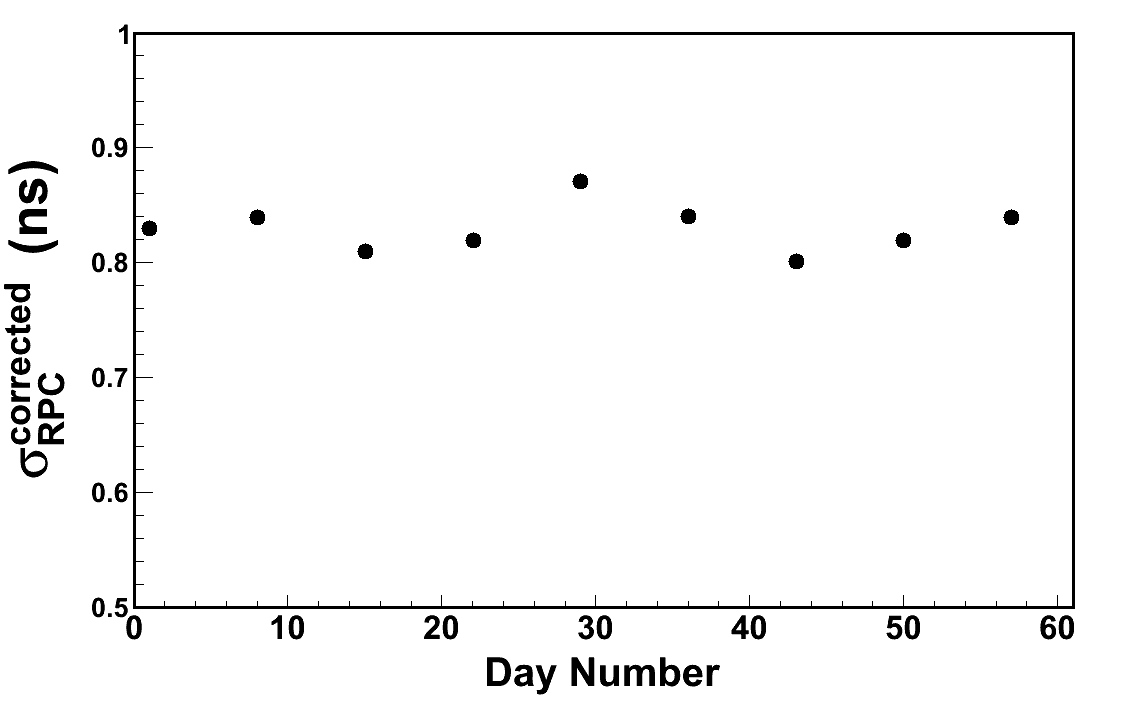}
\caption{\label{time-resolution_time} \small \sl[Color online] Time resolution ($\sigma$) of the chamber over a period of 60 days.}
\label{time-resolution_time}
\end{center}
\end{figure}

\section{ Conclusions}
The RPC has been successfully tested continuously with cosmic rays over a period of 60 days. The chamber remained stable in terms of leakage current,
efficiency, noise rate and time resolution over the entire testing period.
The average leakage current, efficiency, noise rate and time resolution of the RPC tested with cosmic rays in the streamer mode of 
operation at 9000V have been measured to be $\sim$3.0 $\mu$A, $>$95$\%$, $\sim$0.75 Hz/$cm^{2}$ and $\sim$0.83 ns respectively.
The results obtained with this chamber make it suitable to be used for muon detection in large neutrino experiments.

\acknowledgments

The work is partially supported by the DAE-SRC award project fund of S. Chattopadhyay.
We are thankful to the INO collaborators for their encouragement.
We acknowledge the service rendered by Ganesh Das for fabricating the detector. We also
acknowledge Ramanarayana Singaraju for his constant help throughout testing of this detector. 
We take this opportunity to thank Satyajit Saha and Y. P. Viyogi for their constant encouragement and all the fruitful
discussions and suggestions.

\end{document}